\documentclass[english,a4paper]{article}
\usepackage[T1]{fontenc}
\usepackage[cp1250]{inputenc}
\usepackage{amssymb}
\usepackage{hyperref}
\usepackage{amsfonts}
\usepackage{graphicx}
\usepackage{revsymb}
\usepackage{dcolumn}
\usepackage{bm}

\def\openone{\leavevmode\hbox{\small1\kern-3.3pt\normalsize1}}
\usepackage[usenames,dvipsnames]{color}

\begin{document}
\title{On the observation of field-free orientation of a symmetric top molecule by terahertz laser pulses at high temperature}
\author{P. Babilotte, K. Hamraoui, F. Billard, E. Hertz, B. Lavorel\footnote{Laboratoire Interdisciplinaire Carnot de
Bourgogne (ICB), UMR 6303 CNRS-Universit\'e Bourgogne-Franche Comt\'e},\\ O. Faucher\footnote{Laboratoire Interdisciplinaire Carnot de
Bourgogne (ICB), UMR 6303 CNRS-Universit\'e Bourgogne-Franche Comt\'e}, D. Sugny\footnote{Laboratoire Interdisciplinaire Carnot de
Bourgogne (ICB), UMR 6303 CNRS-Universit\'e Bourgogne-Franche Comt\'e, 9 Av. A.
Savary, BP 47 870, F-21078 Dijon Cedex, France and Institute for Advanced Study, Technische Universit\"at M\"unchen, Lichtenbergstrasse 2 a, D-85748 Garching, Germany, dominique.sugny@u-bourgogne.fr}}

\maketitle
%\date{\today}
\begin{abstract}
We investigate experimentally and numerically the field-free orientation of the symmetric top molecule of methyl-iodide at high temperature using a terahertz radiation generated by a plasma induced by a two-color laser beam. The degree of orientation is measured from the free-induction decay emitted by the sample. The observed experimental signal is reproduced with a good accuracy by numerical simulations.
\end{abstract}
%\pacs{33.80.−b, 42.50.Hz, 05.45.−a, 02.30.Ik}

\section{Introduction}
Manipulating the molecular rotational degrees of freedom in gas phase by means of laser fields remains a very attractive topic in quantum control \cite{reviewQC1,reviewQC2} with a wide range of applications in photochemistry extending from chemical reactivity \cite{Warren1993, Stapelfeldt2003} to nanoscale design \cite{Seideman1997,Stapelfeldt1997}, stereochemistry \cite{Rakitzis2004}, surface processing \cite{Seideman1997,reuter2008}, catalysis \cite{Bulthuis1991}, and attosecond molecular dynamics \cite{Krausz2009}. Such phenomena play also a role in quantum computing \cite{shapiro2003} and high-order harmonic generation \cite{Atabek2003,Ramakrishna2007,Ramakrishna2010,houzet2013}. In this setting, molecular alignment and orientation can be identified as crucial prerequisites before exploring more complex control scenarios \cite{reviewseideman,Stapelfeldt2003}. The alignment process \cite{Friedrich1995} is by now a well-established concept from both the experimental and theoretical points of view with recent extensions ranging from the deflection of aligned molecules \cite{gershnabel2010}, the introduction of planar alignment \cite{hoque2011}, the control of molecular unidirectional motion \cite{korech2013,steinitz2014,karras2015}, the study of molecular superrotors \cite{korobenko2016,korobenko2014,milner2015} or the analysis of dissipation effects due to molecular collisions \cite{ramakrishna2005,viellard1,viellard2,milner2014}. The description and the control of molecular orientation are  not currently at the same degree of improvement, in particular from the experimental point of view. Molecular orientation was achieved in the adiabatic regime \cite{Goban2008,Ghafur2009,Filsinger2009,Takei2016,Muramatsu2009,Holmegaard2009}. If no static field is used, a rapid turn-off of the laser field allows to get orientation under field-free conditions. At low temperature, a very high degree of orientation can be obtained using such control strategies and a molecular quantum-state selection \cite{Filsinger2009,Takei2016,Muramatsu2009,Holmegaard2009}. Control schemes in the sudden regime, where the duration of the control field is short with respect to the rotational period, have been also developed \cite{averbukh2001,daems2005,dion2001,dion2005,ortigoso2012,Atabek2003,sugny2005,sugny2014,Lapert2012,machholm2001,henriksen1999,prasad2013,shu2013,Tehini08,zhang,Tehini2012,wu,Kanai2001}.
Among other techniques, we can mention the interaction of the molecule with a terahertz (THz) laser pulse and the $(\omega,2\omega)$- scheme. Note that a larger degree of orientation can be achieved with the two-color  mechanisms through ionization depletion \cite{Spanner2012}. In this second regime, molecular orientation has been recently addressed experimentally for linear molecules using only a THz field \cite{Nelson2011} or its combination with a laser field \cite{Kitano2011}, and in \cite{Znakovskaya2009,Kraus2014} for an excitation process with two-color laser fields.

In this work, we complement the previous experimental and theoretical works on the orientation dynamics produced by THz fields by exploring the orientation at high temperature (typically room temperature) of a symmetric top molecule, CH$_3$I. This molecule is a good candidate to achieve a high degree of orientation at room temperature due to its large permanent dipole moment and its relatively small rotational constant \cite{Lapert2012}. The THz pulses are obtained from the excitation of a plasma by a two-color femtosecond laser field \cite{Cook2000}, while the detection process is based on the free-induction decay (FID) emitted by the molecular sample after the THz excitation \cite{hardprl,hard1991,hard1994,bigourd2008}. We extend the previous studies on the subject by considering the case of a symmetric top molecule at room temperature. We show that a noticeable degree of orientation can be reached. A theoretical description of the propagation of a THz field in the sample shows that the FID is not proportional to the degree of orientation but to its time derivative. A complete analytical derivation of this result is given in this paper. Note that this dependency has been already mentioned in \cite{Nelson2011}. The relation between the FID and the degree of orientation allows us to quantitatively compare the experimental observations with the numerical simulations. A very good match is found for the first two orientation revivals. This agreement has been improved by accounting for the centrifugal distorsion and the relaxation effects in the computations \cite{ramakrishna2005,viellard1,viellard2}. We also use this theoretical description to explore the influence of the laser parameters on the orientation dynamics.

The paper is organized as follows. The experimental setup is described in Sec.~\ref{sec3}, with a special emphasis on the generation of THz pulses and on the detection process. We show in Sec.~\ref{sec2p} that the FID is given at first order by the time derivative of the degree of orientation. The model system is introduced in Sec.~\ref{sec2}. The numerical and the experimental results are discussed in Sec.~\ref{sec4}. Conclusions and prospective views are given in Sec.~\ref{sec5}.

\section{Experimental setup for producing molecular orientation \label{sec3}}
\begin{figure*} [ht]
                \centering
                \includegraphics[width=12cm]{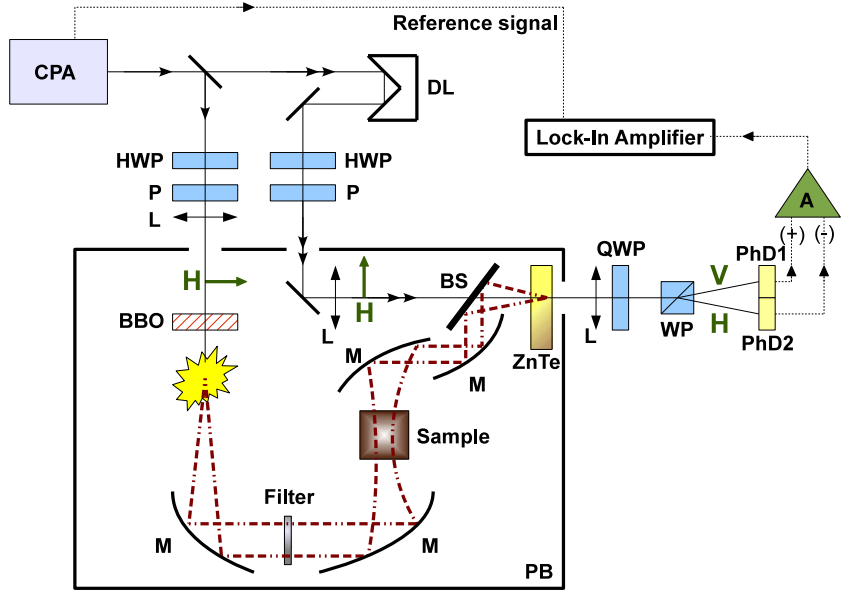}
                \caption{Experimental set-up. CPA: Chirped Pulse Amplifier (100 Hz repetition rate, $\tau_L  \approx 100~\textrm{fs}$, $\lambda_0 = 800~\textrm{nm}$, $E_{100~\textrm{Hz}}  \approx 7.5~\textrm{mJ}$), QWP: Quarter Wave Plate, HWP: Half Wave Plate, P: Polarizer, L: Lens, BBO:  Beta Barium Borate crystal, Filter: see the text, M: off-axis parabolic Mirror, BS: indium tin oxide Beam Splitter, ZnTe: electro-optical crystal, DL: motorized Delay Line stage,  WP: Wollaston Prism, PhD1 and PhD2: balanced photodiodes, A: integrated amplifier, PB: purged box.}
                \label{Set_Up}
\end{figure*}

\begin{figure} [ht]
\centering
\includegraphics[width=10cm]{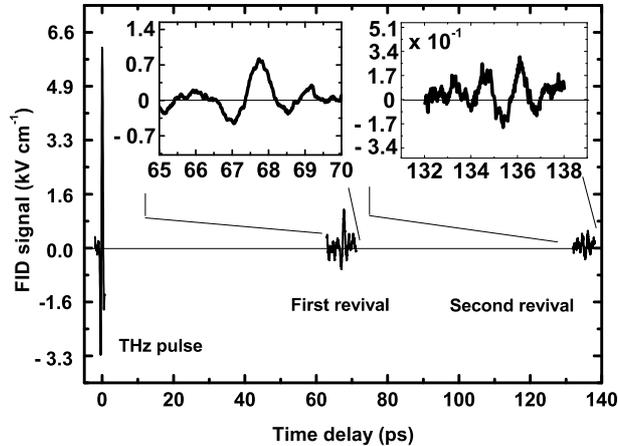}
\caption{Experimental electro-optical sampling signal as a function of the delay between probe and THz pulses in CH$_{3}$I. The pressure in the gas cell was 0.26 bar. The transmitted THz pulse around zero delay and two first orientational revivals at 67 and 134 ps are shown.}
\label{Typical experimental signal}
\end{figure}

The experimental set-up for producing and measuring molecular orientation is shown in Fig.\ref{Set_Up}. It is based on a THz pulsed source and an electro-optical sampling device for the detection. The THz pulses are produced through plasma generation in gases with two-color femtosecond pulses \cite{Cook2000,Kress2004,Kim2007,Kim2008,Vvedenskii2014}. A fraction of the output of a chirped pulse amplifier (CPA, 800 nm wavelength, 100 fs pulse duration, 7.5 mJ energy, 100 Hz repetition rate) is focused in dry nitrogen. A type I phase matching $\beta$-Barium Borate (BBO) crystal  is inserted on the beam path at a given distance from the focusing point to produce the second harmonic at 400 nm. The generation of THz pulses is optimized by adjusting the phase matching angle of the frequency doubling crystal and its longitudinal position using a micrometric stage. The THz source provides pulses of typically few hundred femtoseconds pulse duration with only one cycle (see Fig.~\ref{fig2} below for details) and covers a spectral range from 0 to 4 THz, with a maximum at about 1.5 THz. After collimation by an off-axis parabolic mirror, the two incoming beams at 800 and 400 nm are filtered out by means of a 2-mm thick plate from polytetrafluoroethylene and a thin black polyethylene sheet. The THz pulse is then focused in a cell (40 mm optical pathway) containing the sample at room temperature. The CH$_{3}$I sample (\textit{Sigma Aldrich}, batch reference \emph{BCBK 1300 V}, reagent grade chemicals) is initially stored in the liquid phase and vaporized by expansion into the cell under vacuum, just before the experiment so as to avoid water pollution of the sample.
The THz electric field in the gas sample is less than 100 kVcm$^{-1}$. The typical pressure used in this work is around 0.2-0.3 bar (\textit{e.g.} below the saturation vapour pressure of CH$_{3}$I). For a good transmission of the THz beam through the gas cell, the windows are in polymethylpentene polymer (TPX, 38.1 mm diameter, 2.0 mm  thickness). The transmitted THz radiation is collected from the sample by another off-axis parabolic mirror and sent to the detection device. This latter is based on an electro-optical sampling of the THz pulse shape in the time domain \cite{Winnewisser1997,Cai1998,Gallot1999,Planken2001}. The weak probe beam (typically $\approx 5$ nJ) derived from the output of the CPA is focused and spatially overlapped with the THz radiation in a ZnTe (110) crystal. The polarization of the probe beam is modified through the Pockels effect induced by the THz beam (electro-optic detection process). This allows us to sample the THz electric field by changing the time delay between the two pulses with a motorized delay line stage. The change of the polarization state of the probe beam is measured from the combination of a quarter wave plate (QWP), a Wollaston Prism (WP), and two head-to-tail connected photodiodes so that the difference of their signals is directly obtained  (see Fig.~\ref{Set_Up}). The difference signal is then amplified and sent to a lock-in amplifier synchronized with the laser repetition rate. The quarter wave plate is oriented so as to get a circular polarization without THz pulse and equivalent signals for the two photodiodes. The THz pathway is included in a box continuously purged (relative humidity $\lesssim$ 7 $\%$) with dry nitrogen to avoid absorption by water vapor.
The signal $\Delta{V}$ recorded by the lock-in amplifier is proportional to $\Delta{I}$, the difference between the intensities measured by the two head-to-tail connected photodiodes. $\Delta{I}$ is given by \cite{Planken2001}:
\begin{equation}\label{firsteq}
\Delta{V}\propto\Delta{I}=I_{\textrm{probe}} \omega n^{3} {E(t) } r_{41} \frac{L}{{c}},
\end{equation}
where $I_{\textrm{probe}}$ is the probe intensity, $\omega$ the probe angular frequency, $n$ the refraction index at the probe frequency, $E(t)$ the THz electric field, $r_{41}$ the electro-optic coefficient of the ZnTe crystal (Pockels effect, $r_{41}$=4 pm~V$^{-1}$), $L$ the crystal length (200 $\mu$m), and $c$ the speed of light in vacuum. The typical ratio $\frac{{\Delta V_{\textrm{THz}} }}{{2V_{\textrm{PhD}} }}=\frac{{\Delta{I}}}{I_{\textrm{probe}}}$ measured in the experiment is $\cong$ 1 - 5 $\%$, $\Delta V_{\textrm{THz}}$ being the signal produced by the THz pulse  and $V_{\textrm{PhD}}$ the signal delivered by each photodiode. The electric field $E(t)$ can be directly evaluated by using Eq.~(\ref{firsteq}) and taking into account the transmission and reflection coefficients of the different optical components. The estimated electric field in the gas sample is typically within the range 6~-~30 kV~cm$^{-1}$. Note that $E(t)$ in Eq.~(\ref{firsteq}) is the total electric field including the transmitted THz pulse $E_{0}(x,t)$ and the FID electric field, as discussed below. A typical recording of the electro-optical sampling signal is depicted in  Fig.~\ref{Typical experimental signal}. It exhibits the transmitted THz pulse at zero delay and the two first orientational revivals at 67 and 134 ps. The goal of Sec.~\ref{sec2p} is to interpret this experimental trace in terms of orientation efficiency.
\section{Propagation of a THz field in a gaseous sample \label{sec2p}}
This section is aimed at describing the production of FID and its propagation in a gaseous sample. We follow here the formalism and the approximations used in \cite{hard1991,hard1994,hardprl,bigourd2008}.

To be more concrete, we consider a THz pulse linearly polarized along the $z$- direction and propagating along the $x$- one. This field is of the form $E_{0}(0, t)$ at $x = 0$.
This THz field experiences an instantaneous dipole $d(t)$ along the $z$- direction which satisfies:
\begin{equation}
d(t) = \mu_{0} \langle\cos\theta\rangle(t),
\end{equation}
where $\mu_0$ is the permanent dipole moment, $\mu_0=1.6406$~D for the CH$_{3}$I molecule \cite{gachi1989}, and $\theta$ stands for the angle between the molecular axis defined by the C-I bond and the field polarization direction. The induced dipole $d(t)$ generates a contribution to the THz field which also propagates within the sample.
We introduce the polarization $p(\omega)$ of the sample given in the frequency domain by:
\begin{equation}
p(\omega) =\frac{N}{V}d(\omega)
\end{equation}
where $N$ is the number of molecules and $V$ the corresponding volume. This polarization is also related to the electric field through the susceptibility parameter $\chi(\omega)$:
\begin{equation}
p(\omega) = \epsilon_{0}\chi(\omega)E(\omega)
\end{equation}
where $\varepsilon_0$ is the vacuum permittivity. The spectral distribution of the field $E(\omega)$ is defined as:
\begin{equation}
E(\omega) = \frac{1}{2\pi} \int_{-\infty}^{+\infty} E_0(0, t) e^{i\omega t} dt,
\end{equation}
which leads to:
\begin{equation}
\chi(\omega) = \frac{Nd(\omega)}{\epsilon_{0}V E(\omega)}.
\end{equation}
In a linear approximation framework where $\chi$ does not depend on the amplitude of the field, the resolution of the Maxwell equations gives that the propagation of the field can be written as follows \cite{hard1991,hard1994}:
\begin{equation}
E(x, t) = \int_{-\infty}^{+\infty} E(\omega)e^{i[k(\omega)x-\omega t]}d\omega,
\end{equation}
with $k(\omega) = n(\omega)\omega/c$, $n^{2} = 1+\chi(\omega)$. In the case of a dilute medium, the complex refractive index is given by $n(\omega) \simeq 1+\frac{\chi(\omega)}{2}$. Expanding the exponential term $\exp[\frac{i\omega x\chi}{2c}]$ in Taylor series up to the second order, we arrive at:
%\begin{equation}
%E(x,t) = \int_{-\infty}^{+\infty} E(\omega)e^{i(\omega\frac{x}{c}-\omega %t)}e^{-\frac{\omega}{c}\frac{\chi''}{2}x}e^{i\frac{\omega}{c}\frac{\chi'}{2}x}d\omega,
%\end{equation}
%which can be expressed as:
\begin{eqnarray*}
E(x, t)
&= \int_{-\infty}^{+\infty} E(\omega)e^{i( \omega \frac{x}{c}-\omega t)}d\omega \\
&+ \int_{-\infty}^{+\infty}E(\omega)e^{i( \omega\frac{x}{c}-\omega t)}i\frac{\omega x}{c}\frac{\chi}{2}d\omega ,
\end{eqnarray*}
which leads to:
\begin{equation}
E(x, t) = E_{0}(x, t) + \int_{-\infty}^{+\infty} E(\omega)e^{i( \omega\frac{x}{c}-\omega t)}i \frac{\omega x}{c} \frac{Nd(\omega)}{2\epsilon_{0}V E(0, \omega)} d\omega
\end{equation}
where $E_{0}(x, t)$ is the initial THz pulse.
We then get:
\begin{equation}
E(x,t) = E_{0}(x,t) - \frac{xN}{2c\epsilon_{0}V}\frac{d}{dt}d(t-\frac{x}{c}) ,
\end{equation}
where we have used the fact that the time derivative of a Fourier transform ($FT$) of a function $f$ is given by: $FT[\frac{d}{dt}f(t)] = -i\omega FT[f(t)]$. The THz electric field can finally be written as follows
\begin{equation}
E(x,t) = E_{0}(x,t) - \alpha(x) \frac{d}{dt} \langle \cos\theta\rangle(t-\frac{x}{c})
\label{eq:A10}
\end{equation}
where $\alpha(x)$ is a positive scalar factor depending upon the propagation coordinate $x$. The second term of the right hand side of Eq. (\ref{eq:A10}) is the FID electric field emitted by the molecules of the sample. This contribution is used in the detection process as described above.

From the experimental point of view, we point out that the FID manifests itself as recurrent THz echos launched by the molecular sample after its interaction with the THz pulse. It originates from transient orientation revivals of molecules inducing, under field-free conditions, a non-zero dipole. Another way of interpreting this phenomenon is to consider that the THz pulse experiences spectral shaping during its propagation. Absorption related to $J\to J+1$  transitions produces periodic holes in the spectrum with a spectral separation in angular frequency $\Delta \omega= 4 \pi c B_e$, with $B_e$ the rotational constant (neglecting the centrifugal distortion).  A periodic modulation in the frequency domain leads to periodic replica in the time domain every $\Delta t=2\pi\Delta \omega$. We emphasize that the first term of Eq.~(\ref{eq:A10}) does not reflect the absorption experienced by the incident pulse because of the use of first order expansion.

\section{Model system \label{sec2}}
We describe in this section the model used in the numerical computations to study the control of molecular orientation of the symmetric top molecule of methyl-iodide CH$_{3}$I. We consider the rotational dynamics of this molecular system, which is assumed to be in its ground vibronic state, in interaction with a linear polarized THz field. The Hamiltonian of the system can be written as:
\begin{equation}
	H(t) = H_{0} + H_{\textrm{int}}(t),
	\label{eq:01}
\end{equation}
where $H_0$ and $H_{\textrm{int}}$ describe respectively the field-free Hamiltonian and the interaction with the laser field.
The Hamiltonian of the molecular system is given by \cite{zare}:
\begin{equation}
	H_{0} = B_{e}J^2 + (A_{e} - B_{e})J_{Z}^{2} - D_{J}J^{4} - D_{JK}J^2J_{Z}^{2} - D_{K}J_{Z}^{4}
	\label{eq:02}
\end{equation}
where $J$ is the angular momentum operator and $J_{Z}$ the component of $J$ along the body-fixed $Z$- axis defined by the C-I bond. The energy eigenvalues $E_{JK}$ of the operator $H_{0}$ in the Wigner basis $|JKM\rangle$ for a prolate symmetric top can be expressed as follows:
\begin{eqnarray}
	E_{JK}
	&= B_{e}J(J+1) + (A_{e} - B_{e})K^{2} - D_{J}J^{2}(J+1)^{2}\nonumber \\
	& - D_{JK}J(J+1)K^{2} - D_{K}K^{4},
\label{eq:03}
\end{eqnarray}
where $A_{e}$ and $B_{e}$ are the rotational constants. The states $|JKM\rangle$ are the eigenstates of the square of the angular momentum operator $J^2$ and of its projections, $J_Z$ and $J_z$ on the body-fixed $Z$- axis and space-fixed $z$- axis, respectively \cite{zare}. The molecular parameters of the CH$_{3}$I molecule used in the numerical computations are given in Tab.~\ref{tab:01} \cite{Carocci1998}.
\begin{table}[!htp]
\centering
\begin{tabular}{p{3cm} p{3cm}}
	\hline
	Parameters & Values in cm$^{-1}$ \\ \hline\hline
	$B_{e}$    & $0.25098$                 \\
	$A_{e}$    & $5.173949$                 \\
	$D_{J}$    & $2.1040012\times 10^{-7}$   \\
	$D_{JK}$   & $3.2944780\times 10^{-6}$   \\
	$D_{K}$    & $8.7632195\times 10^{-5}$   \\ \hline
\end{tabular}
\caption{Values of the rotational and of the centrifugal constants of the CH$_{3}$I molecule used in the numerical computations.}
\label{tab:01}
\end{table}
The interaction between the molecular system and the external electromagnetic field reads:
\begin{equation}
	H_{\textrm{int}}(t) = -\mu_{0} E(t) \cos\theta,
	\label{eq:04}
\end{equation}
where the function $E(t)$ represents here the amplitude of the THz electric field. We neglect in this paper the effect of the polarizability components since the maximum intensity of the electric field remains moderate. The units used are atomic units unless otherwise specified.

At room temperature, the system is described by a density matrix $\rho(t)$ whose dynamics is
governed by the Liouville-von Neumann equation \cite{Shapiro2012}:
\begin{equation}
	i\frac{\partial{\rho(t)}}{\partial{t}} = [H(t),\rho(t)],
	\label{eq:05}
\end{equation}
where the initial condition $\rho(0)$ is given by the canonical density operator at thermal equilibrium
\begin{equation}
	\rho (0) = \frac{1}{Z} \sum_{J=0}^{\infty}\sum_{M,K=-J}^{J}  e^{-E_{JK}/(k_{B}T)} |JKM\rangle \langle JKM|,
	\label{eq:06}
\end{equation}
where $Z = \sum_{J=0}^{\infty}\sum_{M,K=-J}^{J} e^{-E_{JK}/(k_{B}T)}$ is the partition function, with $T$ the temperature fixed to $T=298$~K and $k_{B}$ the Boltzmann constant.

The degree of orientation of the molecular system is given by the expectation value:
\begin{equation}
	\langle\cos\theta\rangle (t) = \textrm{tr}[\rho(t) \cos\theta].
	\label{eq:07}
\end{equation}
Furthermore, in order to simulate more realistic experimental conditions (see Sec.~\ref{sec4}), we add to the model system the dissipative effects due to molecular collisions \cite{Seideman2005,viellard1}. To limit the complexity of the numerical computations, we consider in this paper the effective approach proposed in \cite{Seideman2005} to account for coherence relaxation described by the time $T_2/P$, where $P$ is the pressure of the sample. We approximate the decoherence by an exponential decay such that the final degree of orientation is given by:
\begin{equation}
\langle\cos\theta\rangle (t) = \langle \cos\theta\rangle _{\textrm{ND}} e^{-\frac{tP}{T_{2}}},
\end{equation}
where the non-dissipative orientation, $\langle \cos\theta\rangle _{\textrm{ND}}$, is computed from the Liouville-von Neumann equation (\ref{eq:05}).

\section{Numerical and experimental results \label{sec4}}
In this paragraph, we first investigate theoretically the degree of orientation that can be achieved with a THz laser pulse in the experimental conditions of the set-up. The Hilbert space is spanned by the Wigner's functions $|J,K,M\rangle$, with $0\leq J$, $-J\leq K\leq J$ and $-J\leq M\leq J$. $M$ and $K$ being good quantum numbers, the Hamiltonian of the system only depends on the angle $\theta$. Numerically, we consider a finite dimensional Hilbert space with $J\leq J_{\textrm{max}}$, $J_{\textrm{max}}=90$. From a physical point of view, this reduction can be justified by the fact that the THz excitation only transfers a finite amount of energy to the system, which thus stays in a finite dimensional subspace.

\begin{figure} [ht]
		\centering
		\includegraphics[width=8cm]{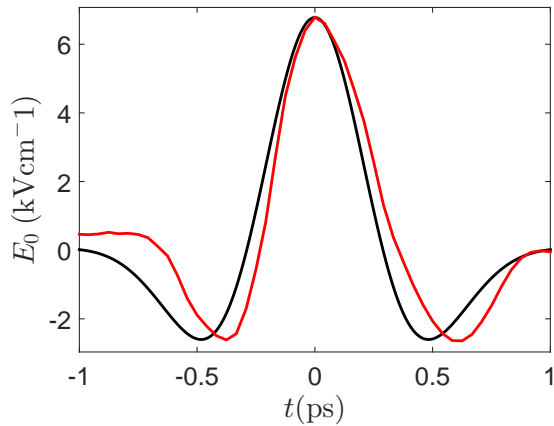}
		\caption{(Color online) Experimental (red or dark gray) and theoretical (black), given by Eq.~(\ref{eqherm}), THz pulses at delay zero. The amplitude $E_1$ has been adjusted to get the best match between the two pulses.}
		\label{fig2}
\end{figure}

The control pulse can be approximated by a set of Hermite polynomials as follows:
\begin{equation}\label{eqherm}
E_0(t)
= \frac{E_1}{2}  e^{ -\frac{t^2}{2\sigma^2}} [ -3 D_{3} \hat{H}_{2}(\frac{t}{\sigma}) + D_{1} \hat{H}_{0}(\frac{t}{\sigma})]
\end{equation}
where $D_{n}=(2^{n}n!\pi^{1/2})^{-\frac{1}{2}}$, and $\hat{H}_{n}(t)$ stands for the Hermite polynomials of order $n$, $\hat{H}_2(t)=4 t^2-2$ and $\hat{H}_0(t)=1$. %Note that the analytical expression  ensures that the pulse has a zero area \cite{sugny2014}.
The parameter $\sigma$ is given by the relation $16\log(2)\sigma^2 = \tau^2$ with $\tau = 1$~ps. The reasonable match between the theoretical THz pulse and the experimental one shown in Fig.~\ref{fig2} justifies the choice made for the electric field in Eq.~(\ref{eqherm}). Note that the peak-to-peak amplitude is experimentally of 9.4~kV~cm$^{-1}$ in this case.

\begin{figure} [ht]
		\centering
		\includegraphics[width=8cm]{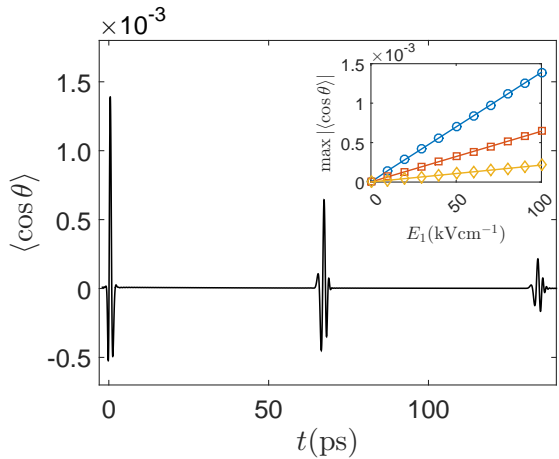}
		\caption{(Color online) Numerical orientation dynamics of the CH$_{3}$I molecule after an excitation with a $1~\textrm{ps}$ pulse centered at $t=0~\textrm{ps}$. The small insert displays the numerical evolution of the maximum degree of orientation at delay zero (blue, circle) and for the first (red, rectangle) and second (yellow, diamond) revivals as a function of the amplitude $E_1$ of the THz field.}\label{fig3}
\end{figure}

\begin{figure} [ht]
		\centering
		\includegraphics[width=8cm]{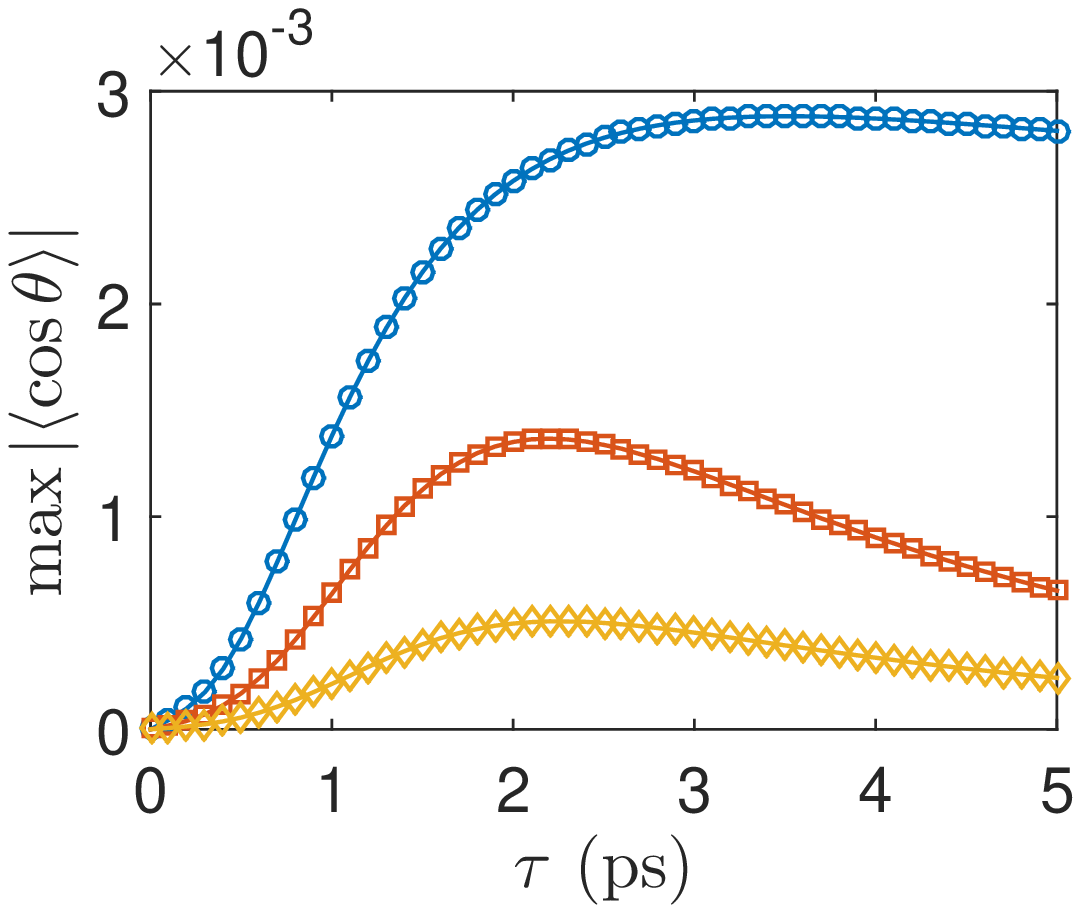}
		\caption{(Color online) Maximum of $|\langle\cos\theta\rangle |$ for the delay zero (blue-circle) and the first (orange-square) and the second (yellow-diamond) revivals as a function of the parameter $\tau$ of the pulse given by Eq.~(\ref{eqherm}). In the experiment, the parameter $\tau$ is of the order of 1 ps.}\label{fig4new}
\end{figure}

We start the analysis of the dynamics by giving a global picture of the time evolution of the molecular orientation as displayed in Fig.~\ref{fig3}, where the relaxation effects are taken into account with $T_{2}= 23$~ps~atm \cite{Hennequin1987,Roberts1968}. The pressure of the cell is $P=0.35~\textrm{bar}$. Here, we fix the amplitude $E_1$ to 100 kV~cm$^{-1}$, which is the maximum experimental available amplitude. Revivals are observed at times multiples of the rotational period. The maximum of orientation is respectively of the order of $5\times 10^{-4}$ and $2\times 10^{-4}$ for the first and second revivals. Note that this maximum is larger than $10^{-3}$ in the non-dissipative case.
In Fig.~\ref{fig3}, we also study how the amplitude of the initial THz field affects the behavior of the orientation dynamics. Due to the low intensity of the field, we observe that the maximum values of each transient evolve linearly with respect to the amplitude. The effect of the pulse duration on the degree of orientation is displayed in Fig.~\ref{fig4new}. More precisely, we consider the role of the parameter $\tau$ as defined in Eq.~(\ref{eqherm}). Note that the overall structure of the field is not modified by this parameter, the field is only compressed or extended in time. This modification corresponds to the currently available shaping techniques of THz pulses. In Fig.~\ref{fig4new}, we observe a nonlinear behavior of the degree of orientation, which is maximum for the two revivals for a value of $\tau$ of the order of 2 ps.

\begin{figure} [ht]
		\centering
		\includegraphics[width=8cm]{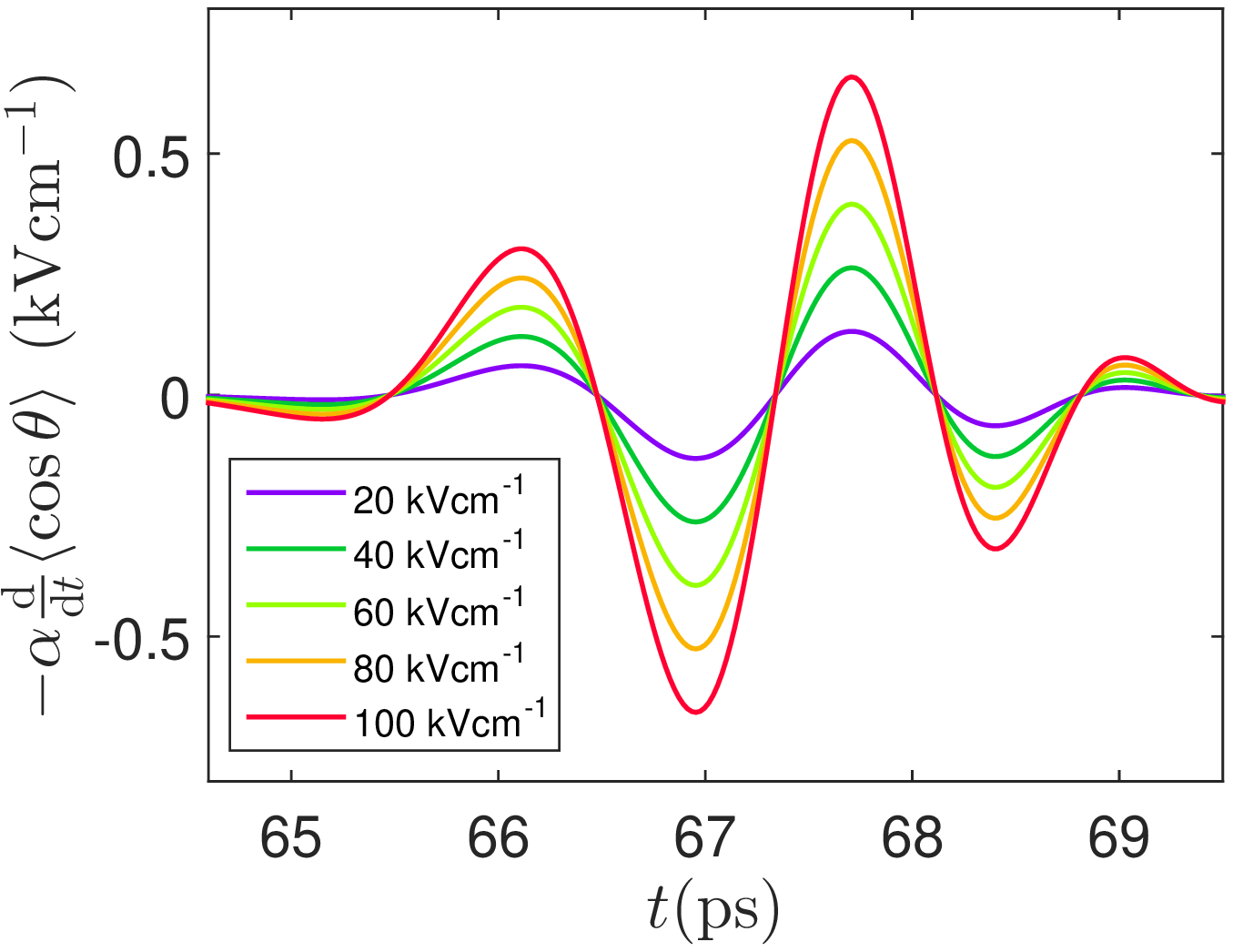}
		\caption{(Color online) Numerical evolution of the time derivative of $\langle \cos\theta\rangle$ of the first revival as a function of the peak amplitude of the electric field which is set to 20, 40, 60, 80 and 100 kV~cm$^{-1}$. The $\alpha$ parameter is set to 1~kV~cm$^{-1}$~ps.}
		\label{var_1tr}
\end{figure}

As shown in Sec.~\ref{sec2p}, the detection process is sensitive to the time derivative of the degree of orientation, i.e. $\frac{d[\langle\cos\theta\rangle]}{dt}$ and not directly to the orientation. In Fig.~\ref{var_1tr}, we study the effect of the amplitude of the field on the first revival. We observe that the shape of the transient does not change when the amplitude is varied. Similar results are obtained for the second revival.

\begin{figure}[htp]
\centering
\includegraphics[scale=0.6]{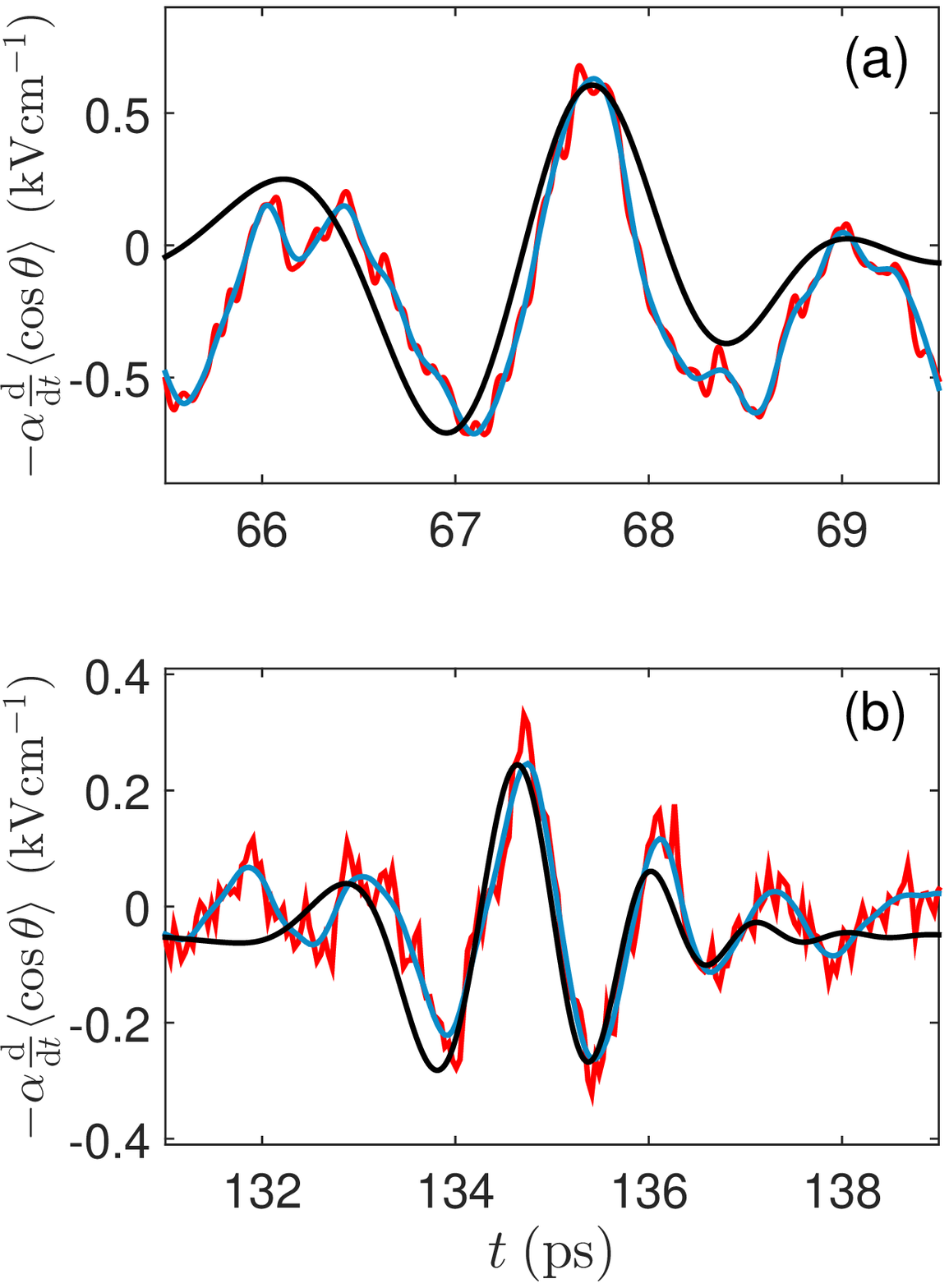}
\caption{(Color online) Time derivative of the degree of molecular orientation of the (a) first and (b) second revivals. The red lines (light gray) represent the experimental data while the black ones correspond to the numerical results. A filtered experimental signal is plotted in blue (dark gray) to ease the comparison with the simulated dynamics. The parameter $\alpha$ is set to 5.75~kV~cm$^{-1}$~ps.}
\label{fig:Figs}
\end{figure}

After this complete theoretical description of the orientation dynamics, the goal is now to make a full comparison of the experimental and theoretical results. A scaling factor and a shift parameter along the vertical axis are determined to get the best match between the two sets of data. As shown in Fig.~\ref{fig:Figs}, the numerical simulation reproduces quite well the experimental signal for the two first transients. The signal is due to a large number of molecules within the volume of the sample, each molecule being excited by a field of different amplitude. The fact that the shape of the revivals does not depend on the intensity of the electric field explains the good agreement observed in Fig.~\ref{fig:Figs}. The signal is too weak after the second revival to pursue this comparison.

\section{Conclusion \label{sec5}}
In this article we have investigated the orientation of a symmetric top molecule, namely CH$_{3}$I. We have shown the efficiency of the full-optical ultrafast set-up resolved in time to generate THz pulses and to produce molecular orientation. We provide a detailed description of the experimental set-up used and a complete numerical study of the corresponding dynamics. The analysis of the detection process shows that the degree of orientation is indirectly measured via the time derivative of the expectation value of $\cos\theta$.  The theoretical model reproduces accurately the experimental results up to the second orientation revival. Additional numerical simulations reveal that the orientation dynamics induced by this THz pulse is qualitatively similar for the linear molecule, OCS. It will be interesting to consider also asymmetric top molecules which have a more complex and non-periodic field-free evolution than linear or symmetric top molecules.

The results of this work can be viewed as an important step forward for the control of molecular orientation. The good match between theory and experiment will allow us to explore the efficiency of more complex strategies using for instance a pre-alignment by a laser field. Such approaches are necessary to increase the degree of orientation and reach efficiencies where molecular orientation could be useful in practice.\\ \\

%\appendix

\noindent\textbf{ACKNOWLEDGMENT}\\
D. Sugny acknowledges the support from the ANR-DFG research programs Explosys (ANR-14-CE35-0013-01) and Coqs (ANR-15-CE30-0023-01). This work was supported by the Conseil R\'egional de Bourgogne under the Photcom Pari program as well as the Labex ACTION program (ANR-11-LABX-01-01) and the CoConiCs program (Contract No. ANR-13-BS08-0013). This work has been done with the support of the Technical University of Munich, Institute for Advanced
Study, funded by the German Excellence Initiative and the European Union Seventh Framework Programme under grant agreement 291763.

%\bibliographystyle{apsrev}
%\bibliography{biblio}% Produces the bibliography via BibTeX.

\end{document}